\documentclass{PoS}

\title{Higgs effective $Hl_il_j$ vertex from heavy $\nu_R$ and applications to LFV phenomenology}
 
\ShortTitle{LFV effective $Hl_il_j$ vertex}

\author{\speaker{Mar\'ia J. Herrero}\thanks{Preprint numbers:IFT-UAM/CSIC-17-092;FTUAM-17-19; LPT-Orsay-17-44. We thank our respective projects: FPA2016-78645-P (MINECO(Spain)/FEDER(EU)); ITN-ELUSIVES H2020-MSCA-ITN-2015//674896 (EU); ANPCyT PICT 2013-2266 (Argentina).}\\
Departamento de F\'{\i}sica Te\'orica and Instituto de F\'{\i}sica Te\'orica, IFT-UAM/CSIC,\\
Universidad Aut\'onoma de Madrid, Cantoblanco, 28049 Madrid, Spain\\
E-mail: \email{maria.herrero@uam.es}}
\author{Ernesto Arganda\\
        Departamento de F\'isica, Universidad Nacional de La Plata,IFLP, CONICET,\\
C.C. 67, 1900 La Plata, Argentina\\
        E-mail: \email{ernesto.arganda@fisica.unlp.edu.ar}}
\author{Xabier Marcano \\
Laboratoire de Physique Th\'eorique, CNRS, \\
Univ. Paris-Sud, Universit\'e Paris-Saclay, 91405 Orsay, France \\
        E-mail: \email{xabier.marcano@th.u-psud.fr}}
\author{Roberto Morales\\
        Departamento de F\'isica, Universidad Nacional de La Plata,IFLP, CONICET,\\
C.C. 67, 1900 La Plata, Argentina\\
        E-mail: \email{roberto.morales@fisica.unlp.edu.ar}}
\author{Alejandro Szynkman \\
        Departamento de F\'isica, Universidad Nacional de La Plata,IFLP, CONICET,\\
C.C. 67, 1900 La Plata, Argentina\\
        E-mail: \email{szynkman@fisica.unlp.edu.ar}}

\abstract{We present a new computation of the Lepton Flavor Violating effective vertex involving the Higgs boson and two leptons with different flavors. This vertex is generated from the integration to one-loop level of the heavy right handed neutrinos which are considered here within the context of the Low Scale Seesaw Models and with masses close to the TeV scale.  We apply the Mass Insertion Approximation technique to compute the loop contributions from these heavy $\nu_R$ and derive a symple analytical formula for the $Hl_il_j$ effective vertex in terms of the input $Y_\nu$ Yukawa coupling matrix and right handed $M_R$ neutrino masses. Some interesting phenomenological applications of this $Hl_il_j$ effective vertex are also included.}

\FullConference{The European Physical Society Conference on High Energy Physics\\
		5-12 July, 2017\\
		Venice}

\begin{document}
\section{Introduction} 
Physical processes involving Lepton Flavor Violation (LFV) in the charged leptonic sector have not been found yet in Nature, but they are intensely searched for at present experiments. The great interest for these searches is obvious: A positive signal of any of these kind of LFV processes would indicate undoubtly the existence of new physics beyond the Standard Model (SM), and most probably would point towards a connection with the origin of neutrino masses and neutrino flavor mixings. Understanding this connection is precisely our major motivation. 
 
\begin{figure}[b!]
\begin{center}
\includegraphics[width=0.6\textwidth]{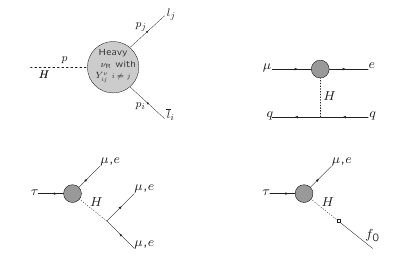}
\caption{Schematic LFV effective vertex $Hl_il_j$ generated from the integration of heavy $\nu_R$ in the loops (shadowed blob), and examples of this vertex participating in Higgs mediated processes: 1) LFV Higgs decays (upper left panel), 2) $\mu-e$ conversion in heavy nuclei (upper right panel), 3) LFV tau leptonic decays (lower left panel), 4) LFV tau semileptonic decays (lower right panel)} 
\label{fig1}
\end{center}
\end{figure}

Our main focus here will be on the practical computation of the $Hl_il_j$ effective vertex that is generated from the integration to one-loop of the heavy right handed neutrinos, within the context of the Low Scale Seesaw Models. Concretely, we will choose to work in the particular realization of these models provided by the Inverse Seesaw Model (ISS), and will assume that the mass scale associated to the right-handed neutrinos $M_R$ is close to the TeV scale, therefore potentially reacheable at the LHC. The other important assumption here is that the charged LFV does not happen at the tree level, but it is radiatively induced to one-loop level from exclusively the heavy $\nu_R$, due to the existence of non-vanishing entries in the neutrino Yukawa coupling matrix, $Y^\nu_{ij}$, with $i \neq j$. These represent the interactions between the Higgs boson $H$, a $\nu_{L_i}$ with a given flavor $i$ and a $\nu_{R_j}$  with a different flavor $j$. An schematic view of this radiatively induced effective $Hl_il_j$ vertex and its role in the LFV Higgs decays and in some Higgs mediated processes is shown in fig.\ref{fig1}.  
For this new computation  we apply the Mass Insertion Approximation (MIA) technique and derive symple analytical formulas for this $Hl_il_j$ effective vertex in the two most important cases: 1) the LFV Higgs decays where the Higgs is on-shell, and 2) the LFV Higgs mediated processes where the Higgs is off-shell. Here we just outline the main results and address the reader to our full paper \cite{Arganda:2017vdb} where all the details and technicalities of this study are presented. 
\section{Theoretical framework: The ISS and the MIA}
We work within the framework of Low Scale Sessaw Models, and more concretely we choose the ISS. The SM particle content is extended with 3 pairs of fermionic singlets with opposite lepton numbers, $\nu_{Ri}(L=+1)$ and $X_j(L=-1)$ with $i,j=1,2,3$. The basic assumption of the ISS, in contrast to other Seesaw Models, is that the Lepton Number global symmetry is only violated by the small $\mu_X$ Majorana mass for the singlets $X$, whereas the mass term for the $\nu_{R}$ singlets, given by the $M_R$ mass scale, is Lepton Number preserving. The ISS Lagrangian includes as well the Yukawa interactions among the left-handed neutrinos, $\nu_{L}$, the right-handed neutrinos, $\nu_{R}$, and the Higgs boson particle, $H$, and is given by:    
\begin{equation}
 \label{ISSlagrangian}
 \mathcal{L}_\mathrm{ISS} = - Y^{ij}_\nu \overline{L_{i}} \widetilde{\Phi} \nu_{Rj} - M_R^{ij} \overline{\nu_{Ri}^c} X_j - \frac{1}{2} \mu_{X}^{ij} \overline{X_{i}^c} X_{j} + h.c.\,, 
\end{equation}
where $L$ is the SM lepton 
doublet, $L= (\nu_L\,\, l_L)^T$,  $\widetilde{\Phi}=i\sigma_2\Phi^*$ with $\Phi$ the SM Higgs doublet containing the Higgs particle $H$, and 
$i,j$ are indices in flavor space that run from 1 to 3. Correspondingly, $Y_\nu$, $\mu_{X}$ and $M_R$ are $3\times 3$ matrices. Without loss of generality, one can choose a basis where $M_R$ is flavor diagonal. 
The neutrino mass matrix and its diagonalization are given by:
\begin{equation}
\label{ISSmatrix}
 M_{\mathrm{ISS}}=\left(\begin{array}{c c c} 0 & m_D & 0 \\ m_D^T & 0 & M_R \\ 0 & M_R^T & \mu_X \end{array}\right)\,,\,\,\,\,
 U_\nu^T M_{\rm ISS} U_\nu = {\rm diag}(m_{n_1},\dots,m_{n_9}). 
\end{equation}
It involves the three relevant mass scales, $M_R$, $\mu_X$ and the Dirac mass, $m_D=v Y_\nu$, with $v = 174\,\mathrm{GeV}$, and  
leads to 9 physical neutrino mass eigenstates $n_i$, $i=1,.,9$, which are Majorana fermions, i.e. they are their own antiparticles. The three lightest ones are identified with the experimentaly detected neutrinos, $\nu$, and the six remaining ones are the new heavy neutrinos, $N$. For the range of our interest, with $\mu_X \ll m_D \ll M_R$, the masses of the light neutrinos are typically of the order of
$m_\nu \sim (m_D^2/M_R^2) \mu_X$, and the masses of the heavy neutrinos are of the order of $m_N \sim M_R$. The six heavy neutrinos indeed come into 3 pseudo-Dirac pairs with small mass splittings between the two components in each pair of the order of $\mu_X$, hence, they are considered as quasi-degenerate.   

For the present computation, we use the MIA and we do not work with the neutrino physical basis, $n_i$, but instead with the Electroweak (EW) basis. Therefore, we use $\nu_L$, $\nu_R$ and $X$ and their corresponding  Feynman rules derived from $\mathcal{L}_\mathrm{ISS}$ in terms of $Y_\nu$, $M_R$ and $\mu_X$. Furthermore, we choose as our input neutrino parameters: $Y_\nu$ and $M_R$. For this purpose, we use the $\mu_X$ parametrization proposed in \cite{Arganda:2014dta}, where $\mu_X$ is not an input but it is derived from $m_D$ (i.e. from $Y_\nu$), $M_R$ and the physical light neutrino masses, $m_\nu$, and mixings in $U_{\rm PMNS}$, in order to accommodate the low energy neutrino data,
\begin{equation}
 {\mu_X}={ M_R^T}{ m_D^{-1}} U_{\rm PMNS}^* m_\nu U_{\rm PMNS}^\dagger { m_D^{T^{-1}}}{M_R}.    
\label{muxparam} 
\end{equation}  

The use of the MIA has the advantage of organizing the computation of the one-loop effective vertex $Hl_il_j$ as a series expansion in powers of $Y_\nu$. This is an important point here since, as we have said, the existence of non-diagonal entries $(Y_\nu)_{ij}$ with $i \neq j$ acts as the only seed for the radiative generation of charged LFV. Concretely, the LFV decay 
$H(p_1) \to l_k(-p_2) \bar l_m (p_3)$ can be written in terms of lepton flavor changing form factors $F_{L,R}$ (we omit the $km$ flavor indices in these form factors, for shortness) as:
\begin{equation}
i {\cal M} = -i g \bar{u}_{l_k} (-p_2) (F_L P_L + F_R P_R) v_{l_m}(p_3) \, \end{equation} which then are expressed in the MIA as: 
\begin{equation}
F_{L,R}^{{\rm MIA}} = \left(Y_{\nu} Y_{\nu}^{\dagger} \right)^{km}f_{L,R}^{(Y^{2})} + \left(Y_{\nu} Y_{\nu}^{\dagger} Y_{\nu} Y_{\nu}^{\dagger}\right)^{km}f_{L,R}^{(Y^{4})} \, +....
\label{MIAFF}
\end{equation}
Here, $P_{L,R}$ refer to the usual Left and Right chiral proyectors, respectively.
The leading order (LO) terms are of ${\cal O} (Y_\nu Y_\nu^{\dagger})$, or  ${\cal O} (Y_\nu^2)$ in short, the Next to Leading Order (NLO) terms are of ${\cal O} (Y_\nu Y_\nu^{\dagger}Y_\nu Y_\nu^{\dagger})$, or ${\cal O}(Y_\nu^4)$ in short, etc.The counting of $Y_\nu$ factors in each diagram can be  easily tracked since by using the EW basis they just come from either the neutrino couplings to the Higgs boson,  $(H \nu_{L_i} \nu_{R_j}) =-(i/\sqrt{2})(Y_\nu)_{ij}P_R$, the neutrino couplings to the charged Goldstone boson, $(G^- l_i \nu_{R_j}) =i(Y_\nu)_{ij}P_R $, or the Left-Right neutrino mass insertion $(\nu_{R_j}-\nu_{L_i})= -i v (Y_\nu)_{ij}  P_R$. In addition, we have used what we call the {\it fat propagators} for the $\nu_R$'s, where the full series with internal insertions of $X$ singlets into the propagation of the $\nu_R$'s (each $X-\nu_R$ mass insertion given by $M_R$) are added up, leading to $P_R~ (i p_\mu \gamma^\mu /(p^2-M_{R}^2))~ P_L$. These propagators turn out to be very convenient since $M_R$ appears effectively in the denominator. More  details can be found in \cite{Arganda:2017vdb}.  
\begin{figure}[t!]
\begin{center}
\includegraphics[width=.49\textwidth]
{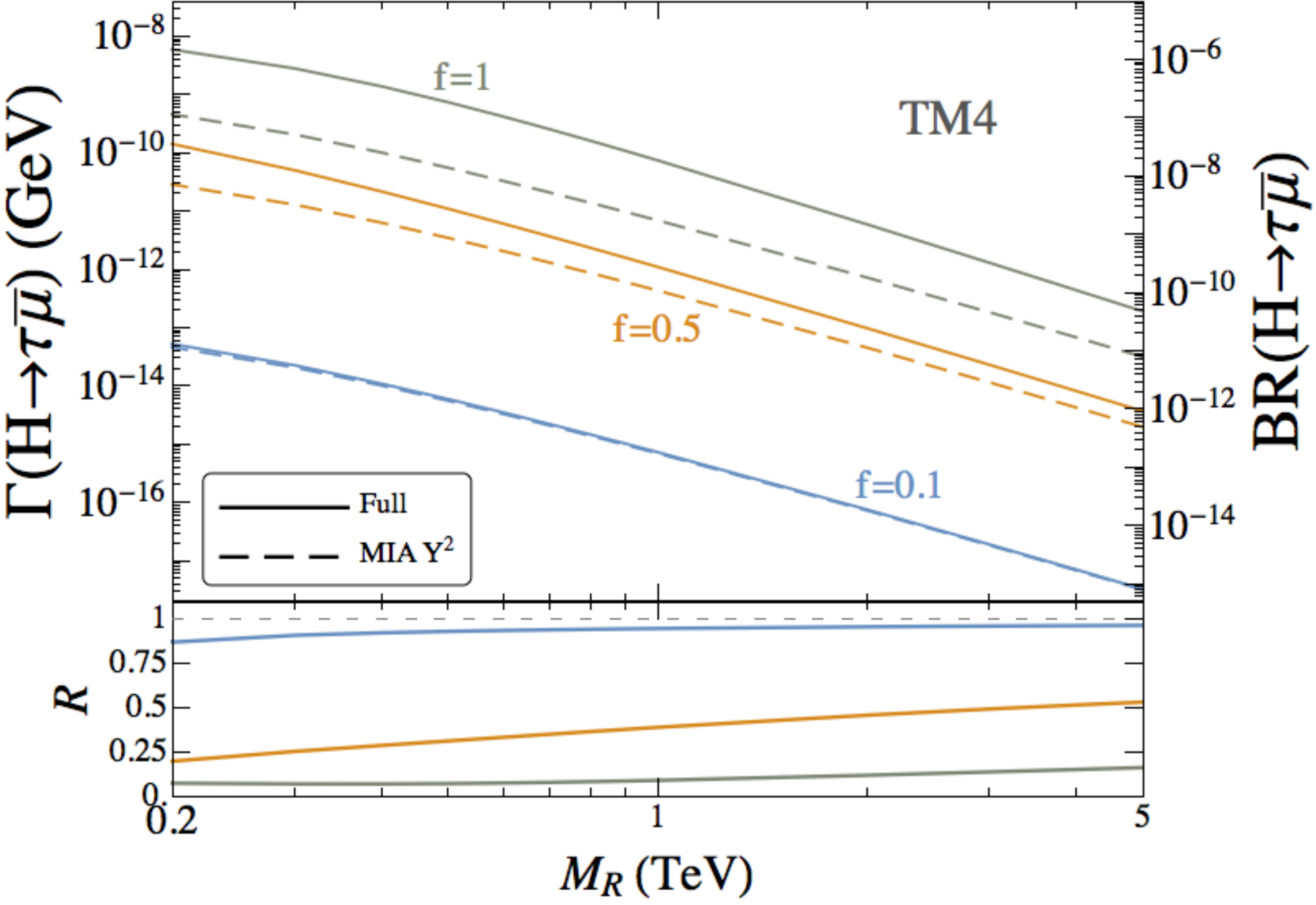}
\includegraphics[width=.49\textwidth]
{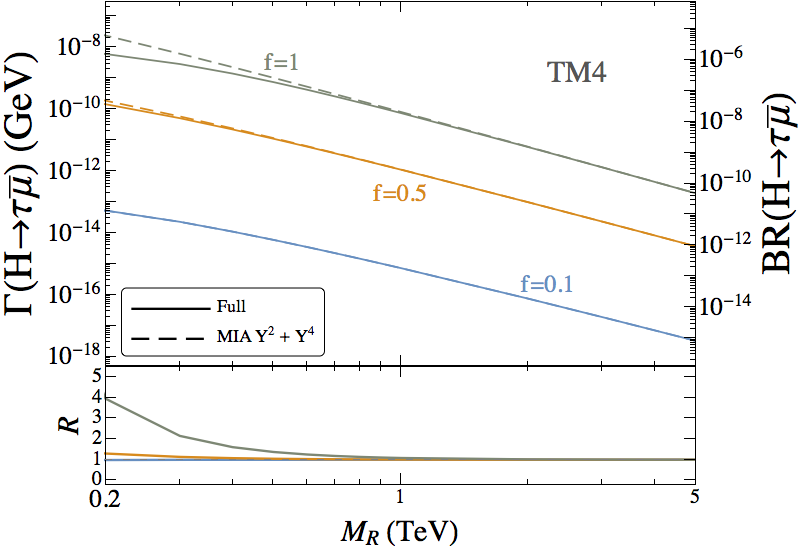}
\caption{Predictions for $H \to \tau \bar \mu$ with the MIA (dashed lines) to leading order, $O(Y_\nu^2)$ (left panel), and to next to leading 
order, $O(Y_\nu^2+Y_\nu^4)$ (right panel), and comparison with the full one-loop result (solid lines). $R$ depicted in the lower part of the plot is the ratio between the MIA and the full results. The input neutrino Yukawa matrix in this plot is (in rows): 
$Y_{\nu}^{\rm TM4}=f \{(0.1 \,\,\, 0\,\,\,0),\,\,\, (0 \,\,\, 1\,\,\,0), \,\,\,  (0 \,\,\, 1\,\,\,0.014)\}$.}
\label{fig2}
\end{center} 
\end{figure}
\section{Results}
To derive the analytical formulas for the MIA form factors $F_{L,R}^{{\rm MIA}}$ above we have performed the explicit diagrammatic computation of all the relevant one-loop diagrams in terms of the standard Veltman-Passarino scalar one-loop functions. In order to show the gauge invariance of our results we have done this computation in: 1) the Feynman-'t Hooft gauge, 2) the unitary gauge and 3) the generic  $R_\xi$ covariant gauges; and we found the same result in the three cases. Specificaly, in the Feynman-'t Hooft gauge the relevant one-loop diagrams are: 25 diagrams of  
${\cal O}(Y_\nu^2)$ and 14 diagrams of ${\cal O}(Y_\nu^4)$) whose drawings and full analytical results can be found in \cite{Arganda:2017vdb}. 

Some selected examples of our numerical results with the MIA for the LFVHD partial widths and their corresponding branching ratios are shown in fig.\ref{fig2}. For comparison with the MIA results, we have also included in these plots the predictions from the full one-loop computation in the physical neutrino mass basis taken from \cite{Arganda:2014dta}. For the numerical evaluation we use the kind of TM and TE scenarios defined in \cite{Arganda:2014dta} which are designed as to get the needed suppression of $\mu-e$ transitions and, correspondingly, the agreement with the most restrictive bounds on these transitions are automaticaly guaranteed. Particularly, this plot is for the scenario TM4, with suppressed $\mu e$ and $\tau e$ but non-suppressed $\tau \mu$ transitions. The results shown in this figure clearly demonstrate that keeping just the LO terms, i.e. ${\cal O}(Y_\nu^2)$, in the MIA expansion, is not sufficient to find a good agreement with the full result, unless the global size of the Yukawa coupling is small, $f \leq 0.1$. Going to the NLO, i.e. ${\cal O}(Y_\nu^2+Y_\nu^4)$, the agreement of the MIA with the full result is very good for $M_R >500$ GeV. This figure also shows clearly the decoupling behaviour of the right handed neutrinos in LFVHD  at large $M_R$.      
 
\begin{figure}[t!]
\begin{center}
\includegraphics[width=.49\textwidth]{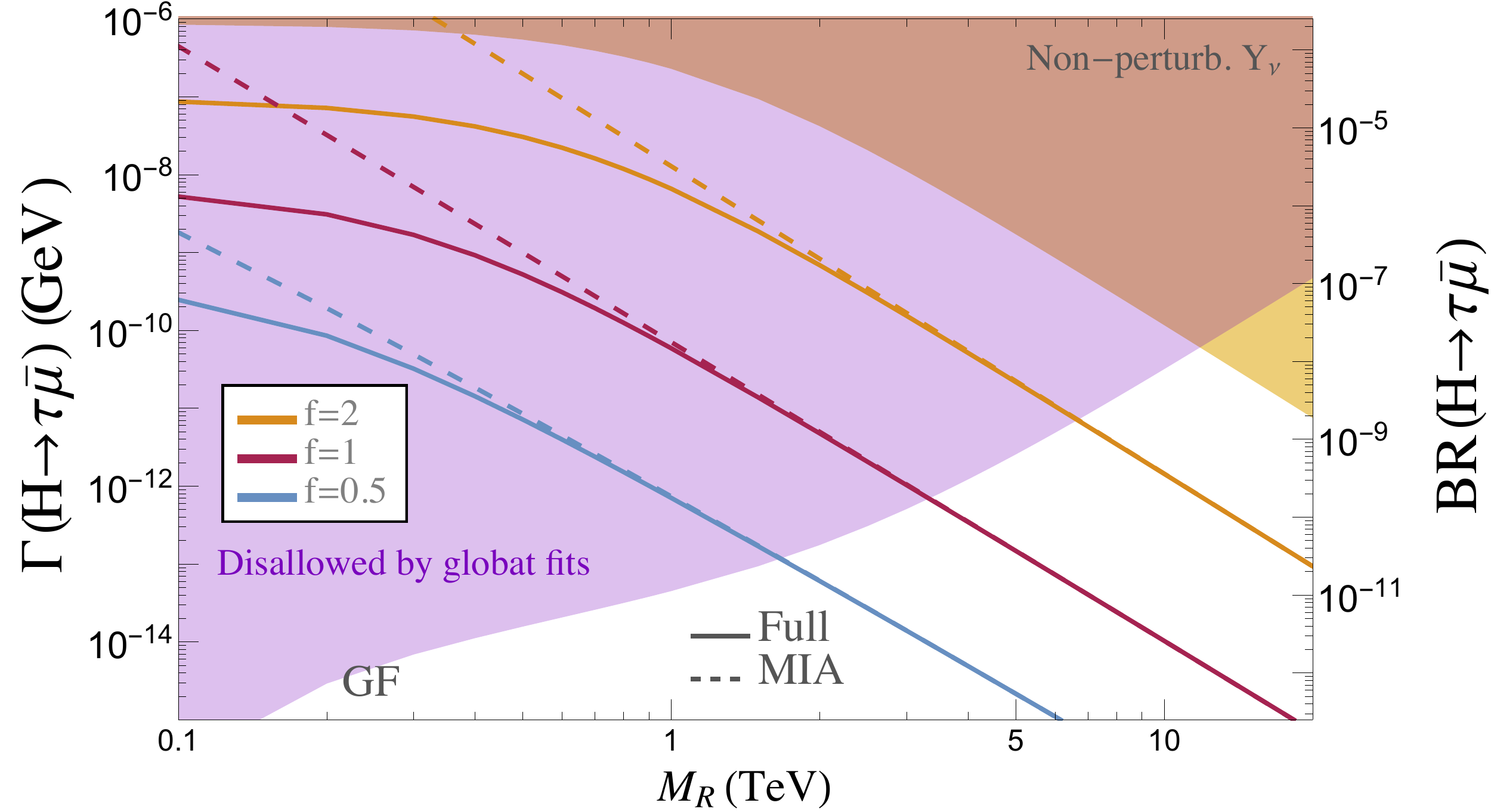}
\includegraphics[width=.49\textwidth]{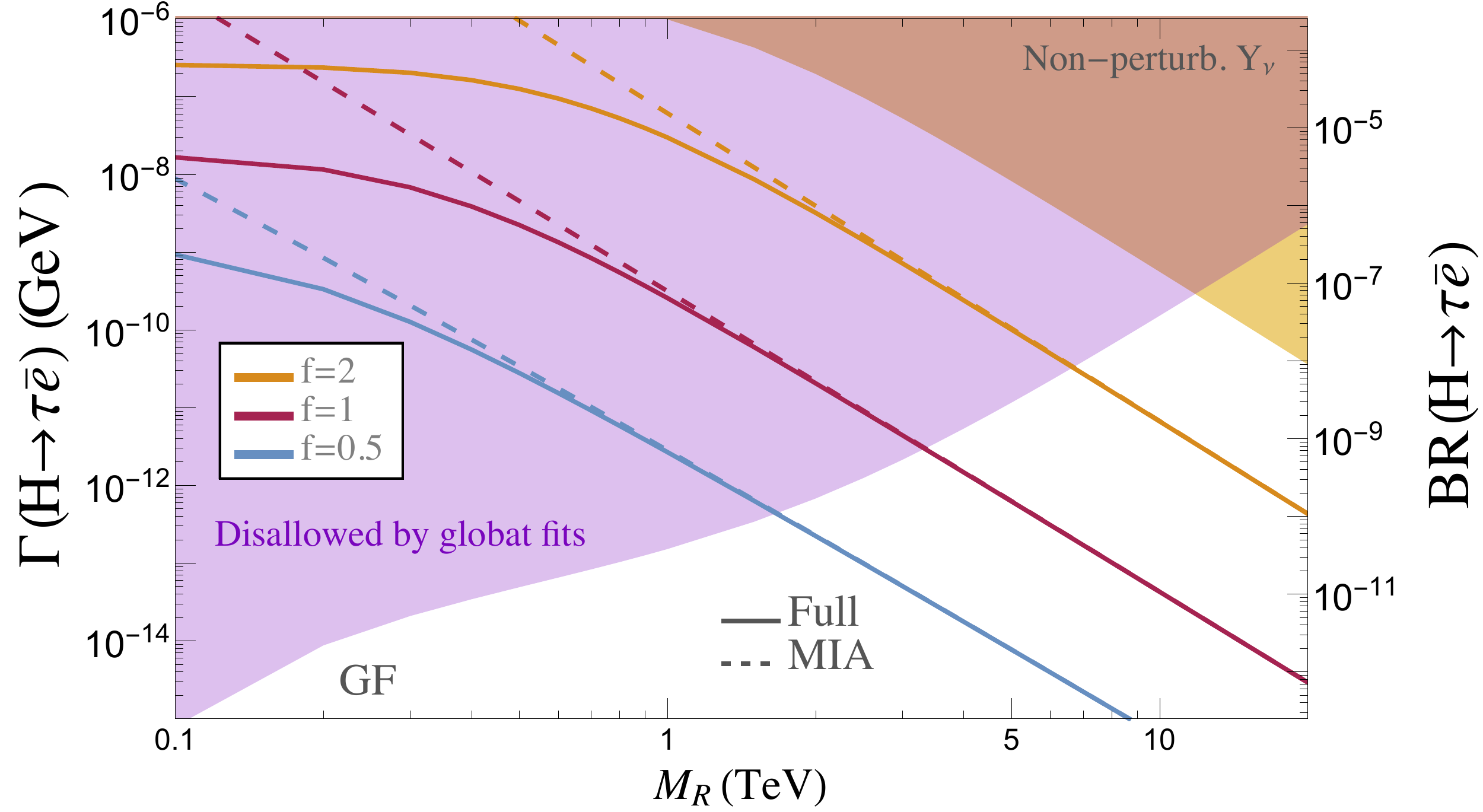}
\caption{Predictions for $H \to \tau \bar \mu$ (left panel) and 
$H \to \tau \bar e$ (right panel) with the effective vertex computed with the MIA (dashed lines). Solid lines are the corresponding predictions from the full one-loop computation in the mass basis. Shadowed areas to the left part of these plots (in purple) are disallowed by global fits. Shadowed areas to the right part of these plots (in yellow) give a nonperturbative Yukawa coupling. The input neutrino Yukawa matrix in this plot is (in rows): 
$Y_{\nu}^{\rm GF}=f \{(0.33 \,\,\, 0.83\,\,\,0.6),\,\,\, (-0.5 \,\,\, 0.13\,\,\,0.1), \,\,\,  (-0.85 \,\,\, 1\,\,\,1)\}$.}
\label{fig3}
\end{center}
\end{figure}

In order to get the final formula for the effective vertex, $Hl_il_j$,  we performed a systematic expansion of the loop integrals in inverse powers of the large mass $M_R$. Our analytical results for the previous functions,  LO $f_{L,R}^{(Y^{2})}$ and NLO $f_{L,R}^{(Y^{4})}$, show explicitly that they are both of order 
$(v/M_R)^2$, whereas the NNLO terms of  ${\cal O}(Y_\nu^6)$ were found to be suppresed as $(v/M_R)^4$, therefore negligible. Assuming heavy $\nu_R$ and  $m_{l_m}\ll m_{l_k} \ll m_W, m_H, m_D \ll M_R$, we found that the dominant contribution goes to the left handed form factor:
 \begin{equation}
i {\cal M} = -i g \bar{u}_{l_k} V_{Hl_{k}l_{m}}^{\rm eff} P_L  v_{l_m} \, ,
\end{equation}
and the simple analytical result for the Higgs on-shell LFV effective vertex reads:
\begin{equation}
V_{Hl_{k}l_{m}}^{\rm eff}=\frac{1}{64 \pi^{2}} \frac{m_{l_k}}{m_{W}}  \left[  \frac{m_{H}^{2}}{M_{R}^{2}}
\left( r(\frac{m_{W}^{2}}{m_{H}^{2}}) +\log(\frac{m_{W}^{2}}{M_{R}^{2}}) \right) \left(Y_{\nu} Y_{\nu}^{\dagger}\right)^{km} - \frac{3v^{2}} {M_{R}^{2}} \left(Y_{\nu} Y_{\nu}^{\dagger} Y_{\nu} Y_{\nu}^{\dagger} \right)^{km} \right]
\end{equation}
where,
 \begin{eqnarray}
r(\lambda)&=&-\frac{1}{2} -\lambda -8\lambda^{2} +2(1-2\lambda +8\lambda^{2})\sqrt{4\lambda-1}\arctan\left(\frac{1}{\sqrt{4\lambda-1}}\right)
\nonumber \\ &+& 16\lambda^{2}(1-2\lambda)\arctan^2\left(\frac{1}{\sqrt{4\lambda-1}}\right) 
\,\,\,;\,\,\, r(m_W^2/m_H^2) \simeq 0.3 \,\,.
\label{rlambda}
\end{eqnarray}
We also obtained the proper off-shell vertex for Higgs-mediated LFV processes, like those in fig.\ref{fig1}, where to take zero external momenta in the effective vertex is a good approximation. We found the simple formula:
\begin{eqnarray}
V_{Hl_{k}l_{m}}^{{\rm eff} \, (p^{{\rm ext}}=0)} = -\frac{1}{32 \pi^{2}} \frac{m_{l_k}}{m_{W}} \left(\frac{3m_{W}^{2}}{2M_{R}^{2}} \right) \left[ \left(Y_{\nu} Y_{\nu}^{\dagger}\right)^{km} + v^{2} \left(Y_{\nu} Y_{\nu}^{\dagger} Y_{\nu} Y_{\nu}^{\dagger} \right)^{km} \right]  \,.
\end{eqnarray}
Finally, we have studied several phenomenological applications of these simple formulas which allow us for very fast and quite accurate estimates of the LFV rates. In particular, the numerical results shown in fig.\ref{fig3} are for $H \to \tau \bar \mu$ and 
$H \to \tau \bar e$ and assume an input neutrino Yukawa coupling matrix $Y_\nu^{\rm GF}$ which is derived by the 'maximum allowed by data' $\eta = (v^2/(2M_R^2))(Y_\nu Y_\nu^\dagger)$ matrix, saturated at the $3\sigma$ level, as extracted from the 'Global Fits' constraints of \cite{Fernandez-Martinez:2016lgt}. Again, the agreement of the MIA and the full predictions is excellent, and we can easily conclude on the maximum LFVHD rates that are allowed by present Global Fits constraints, specifically,  BR$(H \to \tau \bar \mu) \sim 3 \times 10^{-8}$ and BR$(H \to \tau \bar e) \sim 2 \times 10^{-7}$ in these shown examples.  
\section{Conclusions}
We have found very simple and useful formulas for the LFV effective vertex, $Hl_il_j$ in terms of the most relevant neutrino input parameters of Low Scale Seesaw Models, $Y_\nu$ and $M_R$. This vertex collects the main effects from the integration to one-loop level of the heavy right-handed neutrinos and has very interesting applications for LFV phenomenology. In particular, this could also be used by other researchers with their favourite input $Y_\nu$ and $M_R$ settings to get a rapid estimate of the induced LFV rates. 

\end{document}